\documentclass[reprint,amsmath,amssymb,aps,prb]{revtex4-1}

\makeatletter
\def\set@curr@file#1{%
  \begingroup
    \escapechar\m@ne
    \xdef\@curr@file{\expandafter\string\csname #1\endcsname}%
  \endgroup
}
\def\quote@name#1{"\quote@@name#1\@gobble""}
\def\quote@@name#1"{#1\quote@@name}
\def\unquote@name#1{\quote@@name#1\@gobble"}
\makeatother

\usepackage{graphicx}% Include figure files
\usepackage{dcolumn}% Align table columns on decimal point
\usepackage{bm}% bold math
\usepackage{xcolor}

\begin{document}

\title{High frequency mechanical excitation of a silicon nanostring with piezoelectric aluminum nitride layers}

\author{Alessandro Pitanti}
\affiliation{NEST Lab., CNR - Istituto di Nanoscienze and Scuola Normale Superiore, piazza San Silvestro 12, 56217 Pisa - Italy}

\author{Tapani Makkonen}
\affiliation{VTT Technical Research Centre of Finland Ltd, P.O. Box 1000, FI-02044 VTT, Espoo, Finland}

\author{Martin F. Colombano}
\affiliation{Catalan Institute of Nanoscience and Nanotechnology (ICN2), CSIC and BIST, Campus UAB, Bellaterra, 08193 Barcelona, Spain}
\affiliation{Depto. Física, Universidad Autónoma de Barcelona, Bellaterra, 08193 Barcelona, Spain}

\author{Simone Zanotto}
\affiliation{NEST Lab., CNR - Istituto di Nanoscienze and Scuola Normale Superiore, piazza San Silvestro 12, 56217 Pisa - Italy}

\author{Leonardo Vicarelli}
\affiliation{NEST Lab., CNR - Istituto di Nanoscienze and Scuola Normale Superiore, piazza San Silvestro 12, 56217 Pisa - Italy}

\author{Marco Cecchini}
\affiliation{NEST Lab., CNR - Istituto di Nanoscienze and Scuola Normale Superiore, piazza San Silvestro 12, 56217 Pisa - Italy}

\author{Amadeu Griol}
\affiliation{Nanophotonics Technology Center, Universitat Politècnica de Valencia, Spain }

\author{Daniel Navarro-Urrios}
\affiliation{MIND-IN2UB, Departament d'Electrònica, Facultat de Física, Universitat de Barcelona, Martí i
Franquès 1, 08028 Barcelona, Spain}

\author{Clivia Sotomayor-Torres}
\affiliation{Catalan Institute of Nanoscience and Nanotechnology (ICN2), CSIC and BIST, Campus UAB, Bellaterra, 08193 Barcelona, Spain}
\affiliation{Catalan Institute for Research and Advances Studies ICREA, Barcelona, Spain}

\author{Alejandro Martinez}
\affiliation{Nanophotonics Technology Center, Universitat Politècnica de Valencia, Spain }

\author{Jouni Ahopelto}
\affiliation{VTT Technical Research Centre of Finland Ltd, P.O. Box 1000, FI-02044 VTT, Espoo, Finland}

\begin{abstract}
A strong trend for quantum based technologies and applications follows the avenue of combining different platforms to exploit their complementary technological and functional advantages. Micro- and nano-mechanical devices are particularly suitable for hybrid integration due to the easiness of fabrication at multi-scales and their pervasive coupling with electrons and photons. Here, we report on a nanomechanical technological platform where a silicon chip is combined with an aluminum nitride layer. Exploiting the AlN piezoelectricity, Surface Acoustic Waves are injected in the Si layer where the material has been localy patterned and etched to form a suspended nanostring. Characterizing the nanostring vertical displacement induced by the SAW, we found an external excitation peak efficiency in excess of 500 pm/V at 1 GHz mechanical frequency. Exploiting the long term expertise in silicon photonic and electronic devices as well as the SAW robustness and versatility, our technological platform represents a strong candidate for hybrid systems.
\end{abstract}

\maketitle

\section{\label{sec:intro}Introduction}
	
Recent years have witnessed a renaissance in the investigation of micro- and nano-mechanical devices, pushed by the impressive capabilities they gain when combined with photonic and/or electronic systems. Achievements such as zepto and yoctogram mass sensing \cite{Yang2006,Chaste2012,Moser2013}, acceleration sensing as low as 10 $\mu $g$\cdot$Hz$^{-1/2}$ \cite{Krause2012}, low power, fast light intensity \cite{Winger2011,Pitanti2015,Balram2016} and, more recently, polarization \cite{Zanotto2020} modulation are just a few examples of the technology enabled by opto- and electromechanical nanosystems \cite{Aspelmeyer2014}. Following an avenue devoted to fundamental physics, these mescoscopic devices have rightfully entered the realm of quantum mechanics, inaugurated by the milestone achievement of ground state cooling of the motion in micrometric-sized optomechanical \cite{Chan2011} and electromechanical \cite{Teufel2011} devices. As textbook quantum harmonic oscillators, pervasively coupled with many different quantum excitations, nanomechanical systems represent the perfect tool for interfacing different physical objects, from photons to electrons, from excitons to spins. This possibility has been manifested by coherently converting information with high efficiency, with practical demonstration with bi-directional conversion of photons from near-infrared range (NIR) to microwaves (MW) \cite{Andrews2014,Vainsencher2020,Arnold2020}, from NIR to NIR \cite{Hill2012} and from MW to MW \cite{Fink2019}. A particularly intriguing approach would see interfacing NIR and microwave photons, the former being suitable for long-distance telecommunication (thanks to their robustness against thermal decoherence and to low-loss optical fibers) while the latter would be an ideal tool for manipulating and interrogating superconducting qubits, which operates in cryogenic environments. The main challenges of realizing this kind of device lie in the very different footprints of NIR and MW resonators which have to be mutually coupled to the very same mechanical resonator. Moreover, NIR photons suffer from metal-induced Ohmic losses while metal superconducting circuits switch to their normal phase when illuminated with photons with thermal energy above their critical temperature, which in most superconductors lies around 1 K. To keep NIR and MW resonators well separated, a mechanical traveling wave approach is then usually preferred. 
\begin{figure}[t!]
\centering
\includegraphics[width=8cm]{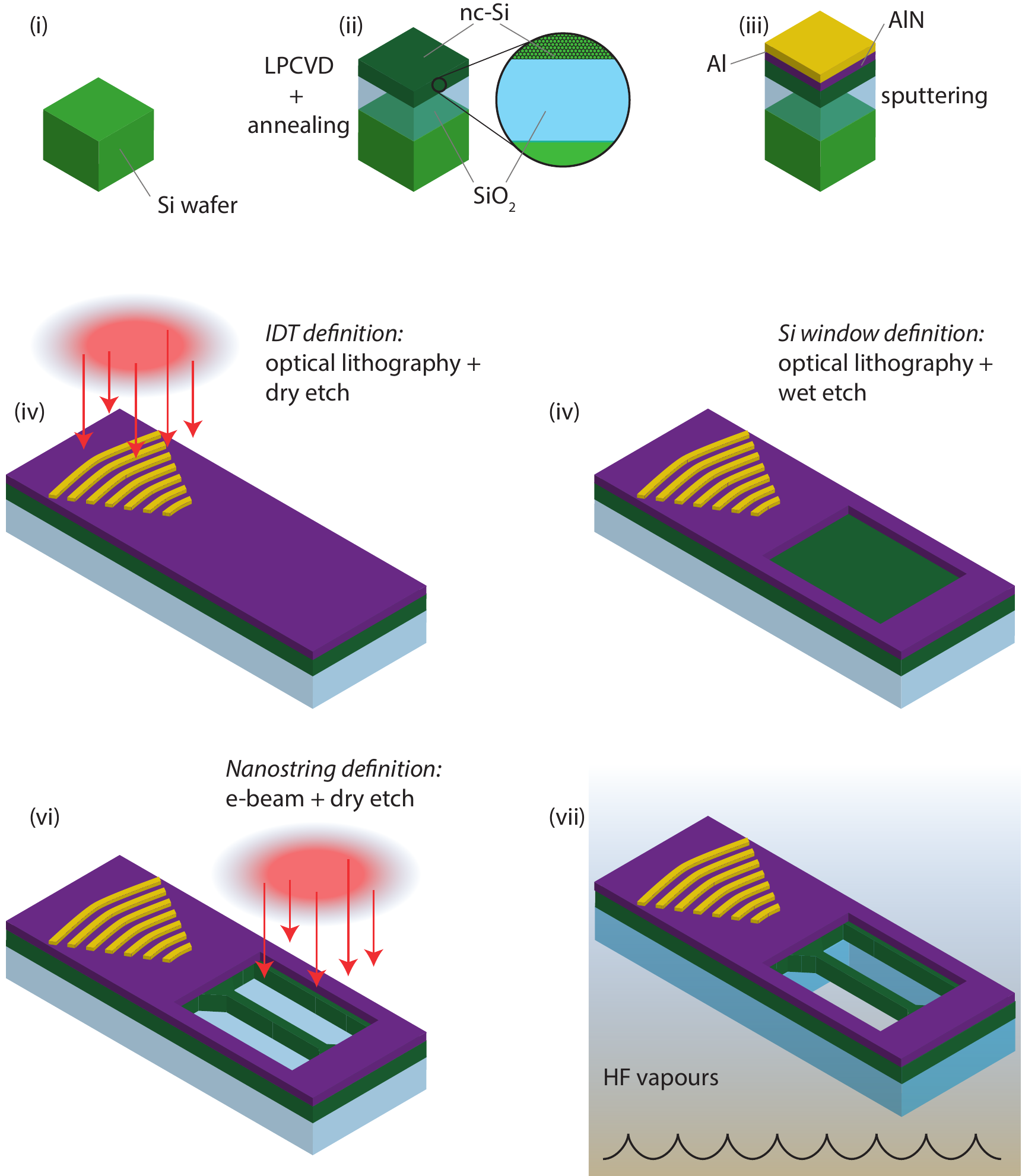}
\caption{Fabrication steps for material growth (i - iii) and device definition (iv - vii).}
\label{fig:1}
\end{figure}
To this end the ideal choice falls in using Surface Acoustic Waves (SAW) which can oscillate at high-frequency ($\sim$ GHz), propagate over long ($\sim$ mm) distances, and have already been used to coupled together qubits separated by 100 $\mu m$ \cite{Gustafsson2014} and pushed in the quantum limit with the generation of non-classical, mechanical Fock states \cite{Satzinger2018}. The SAW is a Rayleigh wave. It does not require piezoelectricity to propagate, but it is easily generated by fabricating metallic InterDigitated Transducers (IDT) on a piezoelectric material. This approach requires a simple AC bias to be applied to the IDTs rather than strong, pulsed lasers which can generate SAW in a non-piezoelectric material through thermally induced local deformations \cite{Hutchins1988}; despite the reduced generation efficiency, the latter approach has been recently applied to Silicon-On-Insulator (SOI) material for optomechanical applications \cite{Munk2019}. Within the electrically-generated SAW scenario, several material platforms have been investigated \cite{Wu2020}, starting from compact monolithic builds made of GaAs \cite{Balram2016,Forsch2020}, AlN \cite{Fan2013} or LiNbO$_3$ \cite{Liang2017}. Another possibility sees the integration of piezoelectric layers on non-piezoelectric materials; combinations such as AlN on Si for mid-infrared photonics \cite{Dong2019} or ZnO on Si/Ge \cite{Boucher2014} have been reported in the literature.\\
In this Article, we demonstrate a new technological platform, which combines the optimum optical properties of Si with a piezoelectric AlN for electrical control of SAW excitation. Despite the discontinuities at the material interface, we show injection of 1 GHz vibrations in a nanostring with external peak efficiency of about 519 pm/V, making this material combination a strong candidate for optomechanical-based wavelength conversion and realization of extended phononic circuits on chip which can be interfaced with electrical or optical systems.

\section{\label{sec:fab}Device Fabrication}
\begin{figure*}[t!]
\centering
\includegraphics[width=16cm]{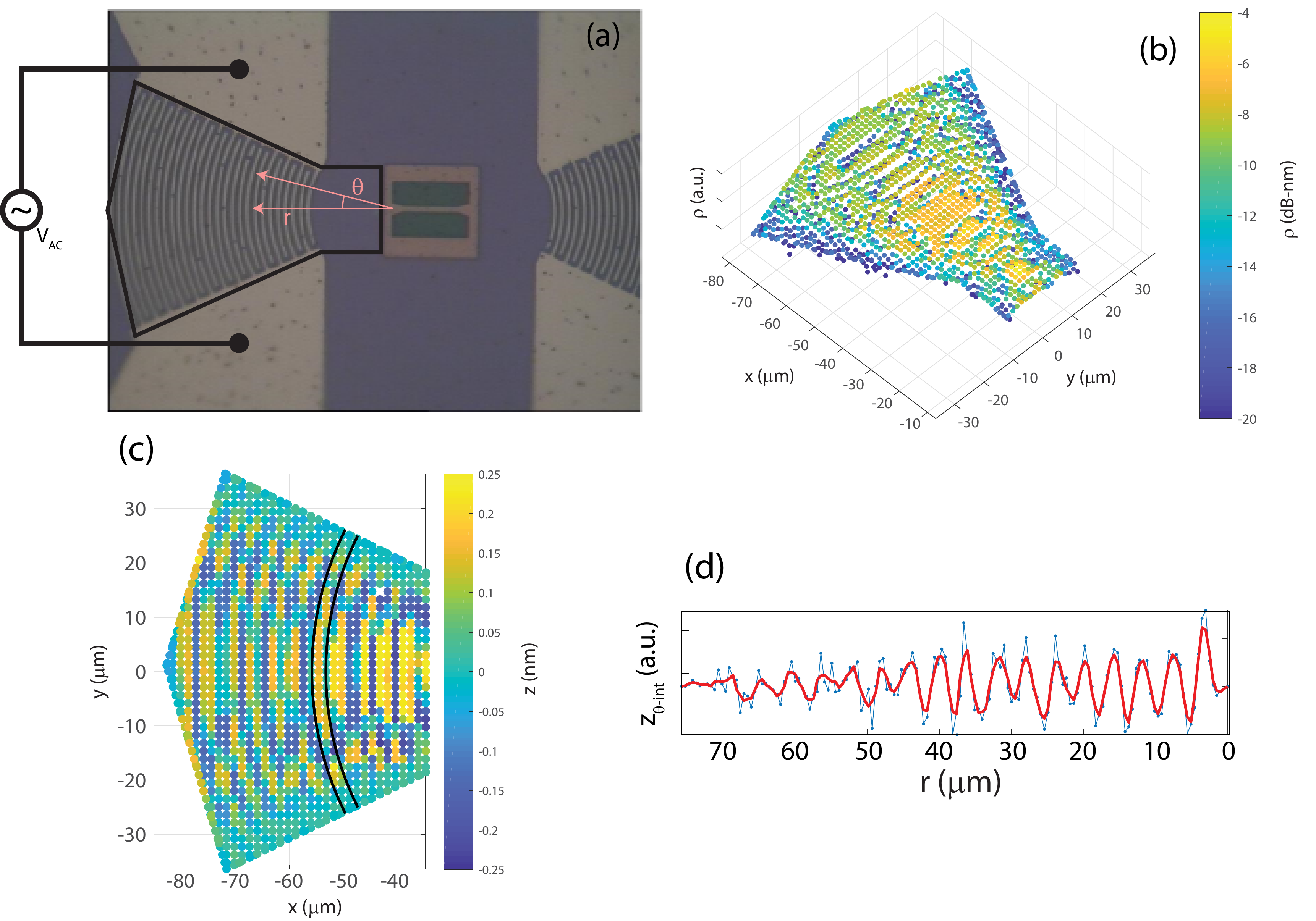}
\caption{a: Microscope image of the device under test. A sketch of the measured region as well as electrical connections are added for clarity as well as the polar coordinate system (in pink) (b): Amplitude map of the selected region for an electrical signal of 2.25 V$_{RMS}$ at 1 GHz. Note that the colorbar is in log-scale. (c): Details of the complex amplitude $z$ in the outer region of the IDT. Integrating along the fingers curvature, the displacement as a function of the geometrical center can be extracted (d), with the red-line a 3-point smoothing. The increase of the amplitude as a result of the IDT focusing effect is clearly seen.}
\label{fig:2}
\end{figure*}
The device fabrication begins with the growth of 1000 nm of wet oxide on a 150 $\mu$m silicon wafer. On the oxide, 220 nm of amorphous silicon is deposited by low pressure CVD at 560 $^\circ$C  (Fig. \ref{fig:1} (i)). The amorphous film is transformed to nanocrystalline by annealing at 950 $^\circ$C for 60 mins\cite{Harbeke1984,Navarro-Urrios2018},  producing a film with randomly oriented single crystalline grains with the size distribution ranging from a few nm to 250 nm (Fig. \ref{fig:1} (ii)). On the annealed nanocrystalline silicon (nc-Si), 250 nm of aluminium nitride (AlN) is sputtered to form the piezoelectric actuator film (Fig. \ref{fig:1} (iii)). Finally, 100 nm of aluminium is deposited by sputtering on the AlN layer for the interdigitated transducers (IDT). For operation at 1 GHz, the IDT finger periodicity is 2 $\mu$m with a 50 $\%$ filling factor. Each finger in a pair is connected to different electrodes, making the system periodicity, upon the feeding of AC bias, of 4 $\mu$m. The total number of finger pairs is 27.
The IDTs are patterned by optical lithography and dry etching (Fig. \ref{fig:1} (iv)). The windows for the nanostrings are selectively opened in the AlN layer using slightly alkaline wet etching Fig. \ref{fig:1} (v)). The nanostrings are patterned by electron beam lithography and dry etching (Fig. \ref{fig:1} (vi)) and, finally, the strings are released in HF vapour, removing the silicon dioxide underneath the strings Fig. \ref{fig:1} (vii)). Amorphous silicon deposited by LPCVD is originally under compressive stress. During annealing, the stress changes to tensile and the amount of stress can be controlled by the annealing conditions. Having a little tensile stress in the device layer is crucial for the fabrication of released structures to avoid buckling and to control the reproducibility of mechanical properties. For the experiments, the devices are wire bonded on a PCB with co-axial connectors for the electrical input.  

\section{\label{sec:char}Device characterization}

The mechanical action of the device has been characterized using a calibrated commercial Laser Doppler Vibrometer (LDV) from Polytech (UHF-1200). The LDV operates in a frequency range up to 1.2 GHz and it is sensitive to mechanical out-of-plane vibrations. The 532 nm solid-state laser is focused on the sample surface using a 100$\times$ objective, resulting in a beam spot of about 1 $\mu$m insize. The full instrument operates as a standard Doppler interferometer with the added possibility of performing digital cross-correlation between the reference signal used to drive the IDT and the output signal as measured by the interferometer. This returns a measure of both quadratures which translates in a complex, time dependent displacement $z(t)=\rho\cdot e^{i(\omega_m + \phi)t}$, with $\rho$ displacement amplitude, $\phi$ displacement phase and $\omega_m$ mechanical frequency. An optical microscope image of the device under test is shown in Fig. \ref{fig:2} (a). The IDT is composed by 13 finger pairs, in shape of circumference arcs of variable radius with the center at the first end of the silicon nanostring. This is done in order to focus the generated SAW directly at the nanostring input. The angular aperture of the arcs is 60$^\circ$. Then device characterization began by centering the laser spot of the interferometer roughly in the center of the IDT fed with a 1 GHz bias. Here we performed a voltage sweep while measuring the displacement amplitude $\rho$. We observed that the device operates in the linear regime up to 2.5 V$_{RMS}$ and therefore we picked 2.25 V$_{RMS}$ as the standard value for the monochromatic device characterization. A 3-dimensional map of the displacement amplitude on top of the IDT is reported in Fig. \ref{fig:2} (b). Here the color bar is in logarithmic scale with the dB-nm defined as $\rho \textrm{[dB-nm]}=10\cdot log_{10}(\rho/1[\textrm{nm}])$. Note that the displacement increases along $\hat{x}$, in the direction closer to the arcs geometrical center as expected from the focusing effect of a curved IDT. 

\begin{figure}[t!]
\centering
\includegraphics[width=8cm]{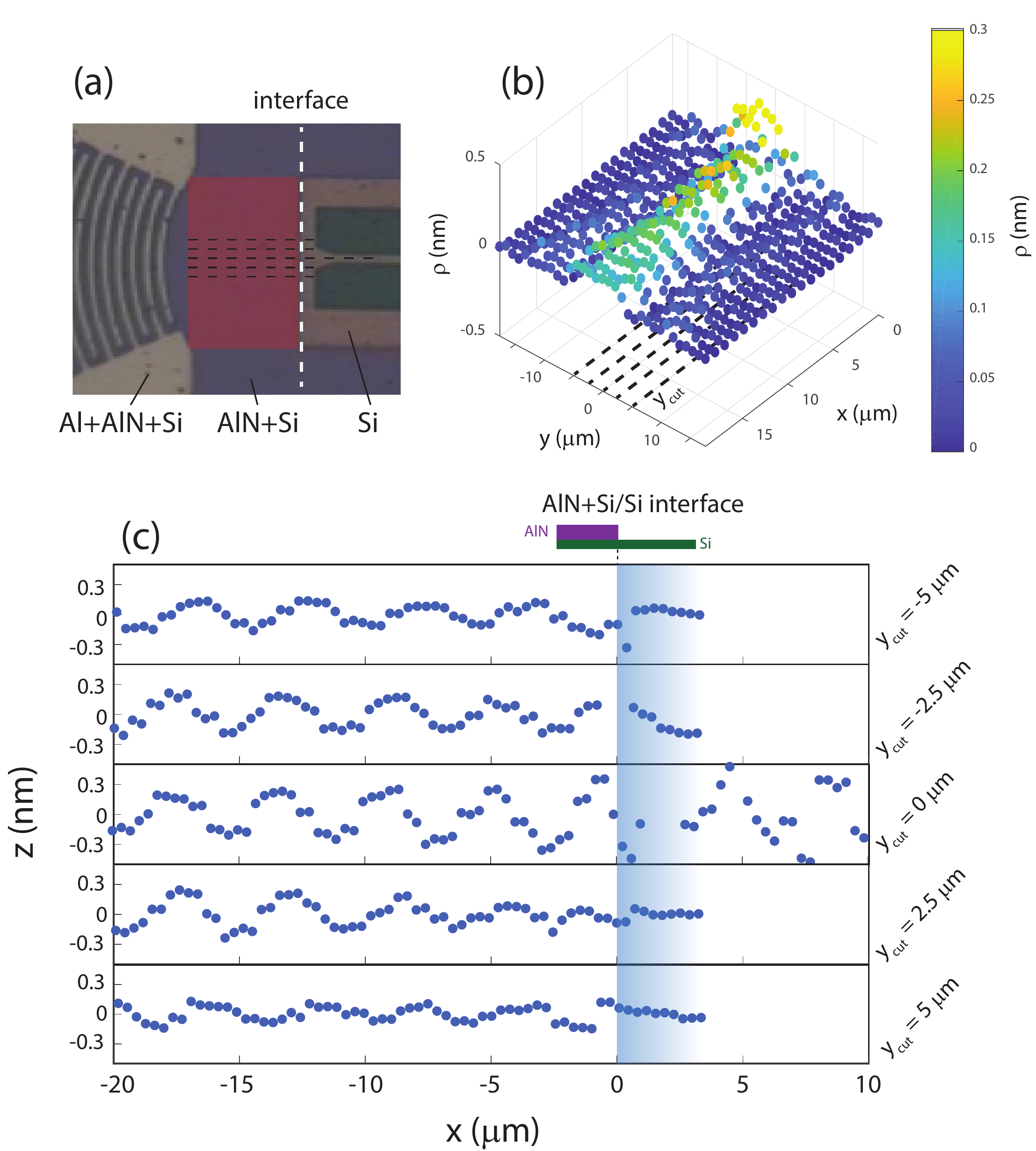}
\caption{(a): Microscope image of the unpatterned focusing region of the IDT. (b): Amplitude displacement map in the red region sketched in (a). (c): Complex displacement at a fixed phase for several line-scans cut at fixed $y$. The interface between the silicon regions with (AlN+Si) and without AlN (Si) is further depicted. Note that the noisy signal at the interface is due to the light scattering from the almost vertical edge.}
\label{fig:3}
\end{figure}

A better view of the wave periodicity on the IDT can be seen by inspecting the z-displacement in Fig. \ref{fig:2} (c). Exploiting the common center of the IDT finger arcs, it is possible to plot the full IDT map in polar coordinates (see the sketch in Fig. \ref{fig:2} (a)) and then integrate along the polar angle $\theta$ from -30$^\circ$ to -30$^\circ$, that is along the black lines sketched as an example in Fig. \ref{fig:2} (c). The results are reported in Fig. \ref{fig:2} (d), which shows a periodic signal with increasing amplitude towards the center. The Fourier transform of this signal has a narrow peak around a frequency of $1/4.02\, \mu m^{-1}$ which very well reproduces the design periodicity of 4 $\mu$m. As mentioned, the signal increases towards $r=0$ due to the focusing effect of the curved IDT. More insights on the focusing effect can be gained by looking at the homogeneous region in front of the last pair of IDT fingers, highlighted in red in the optical microscope image of Fig. \ref{fig:3} (a). A map of the displacement amplitude measured in this region is shown in Fig. \ref{fig:3} (b). 

\begin{figure}[ht!]
\centering
\includegraphics[width=8cm]{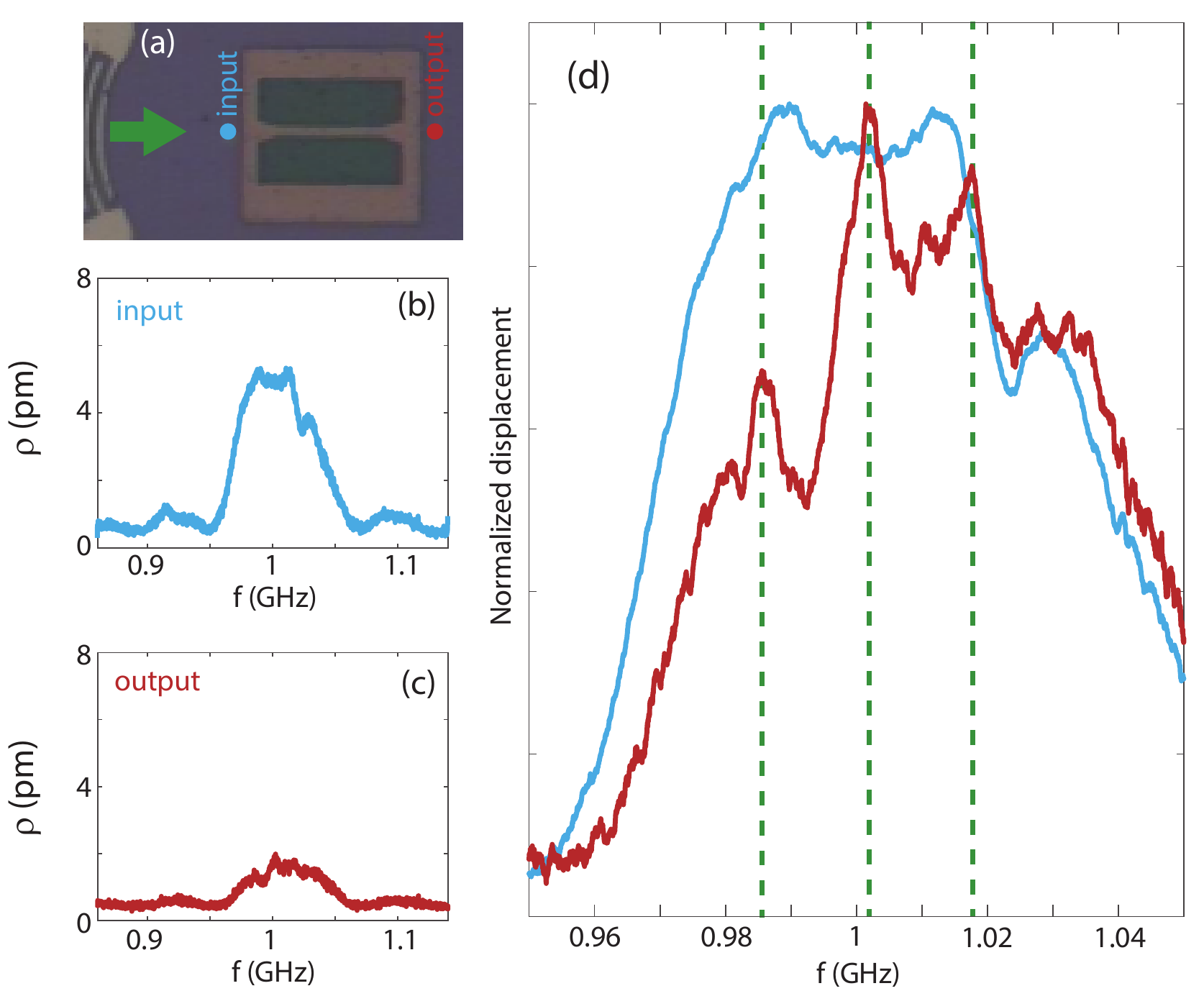}
\caption{(a): Optical microscope image of the nanostring devices. The green arrow indicates the SAW propagation whereas the two colored dots denote the acquisition position for the input - panel (b) and output - panel (c) displacement amplitude spectra. (d): Comparison of normalized input (blue) and output (red) displacement amplitude spectra.}
\label{fig:4}
\end{figure}

\begin{figure}[t!]
\centering
\includegraphics[width=8cm]{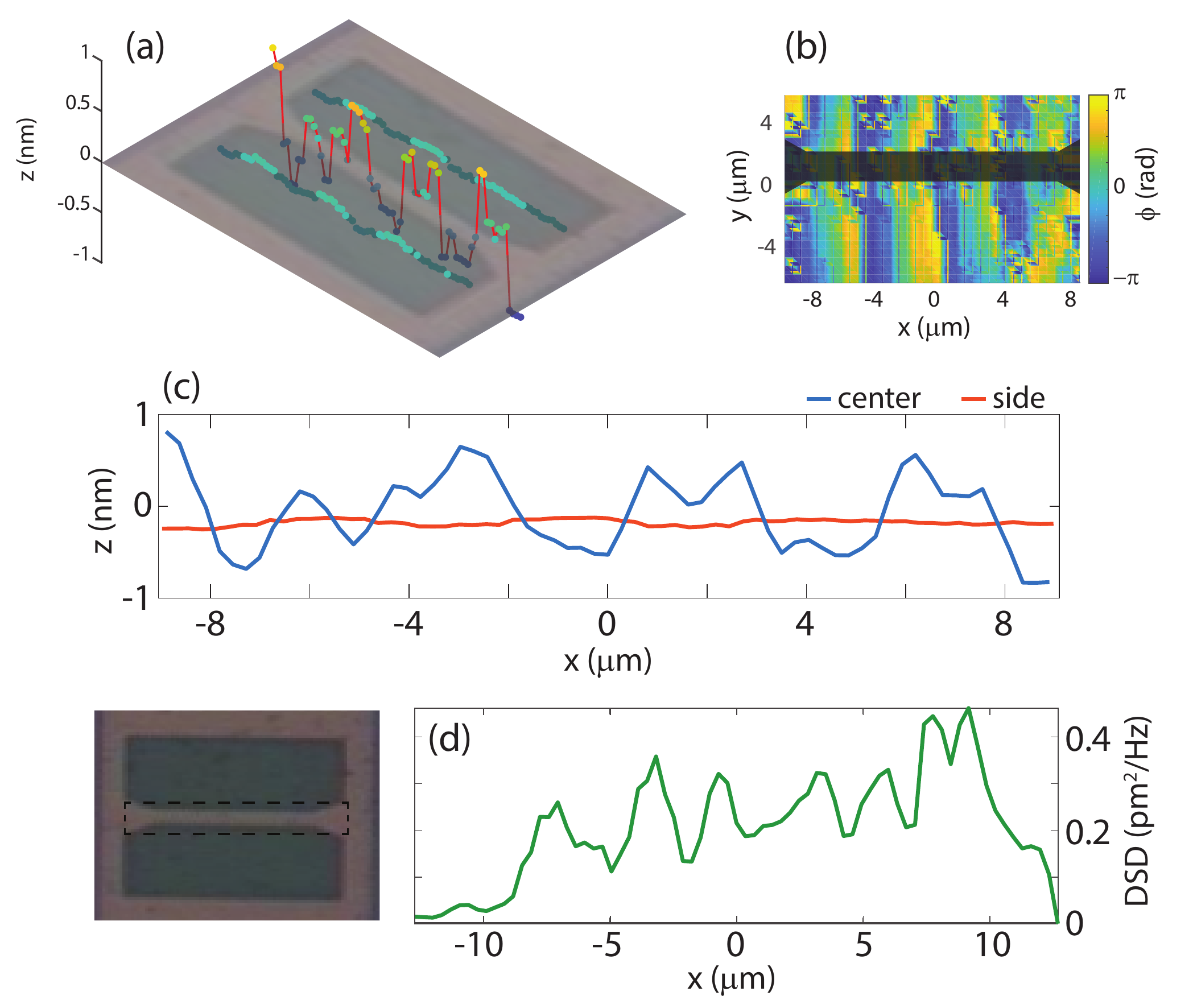}
\caption{(a): 3D sketch of the line scan measurement showing a stronger excitation signal on top of the nanostring with respect to the ones coming from the residual substrate. (b): Phase map of the displacement in the etched region. (c): Direct comparison of the displacement along the central and one side linescans of panel (a). (d): Average DSD along the nanostring cross section}
\label{fig:6}
\end{figure}

The displacement field is spatially narrowing and increasing in amplitude when moving in $\hat{x}$ direction towards the nanostring, reaching a maximum value of about 300 pm. This effect can be better quantified by plotting the complex displacement at a fixed phase (see Fig. \ref{fig:3} (c)) along the dashed black lines sketched in Fig. \ref{fig:3} (a) and (b). A periodic signal can be observed in every line-scan with a strong difference in the signal amplitude along the central cut ($y=0$) as compared to the other ones. In the center, the signal is increasing and propagates strongly even after the AlN+Si/Si interface. Beside the central line, the amplitude decreases slightly along $\hat{x}$, with a net reduction after the interface. Moreover, the absolute value of displacement becomes smaller further away from the central line. The results can be better appreciated considering the material interface which is depicted with a blue-shaded region, while a top view of the different material stacks can be seen in Fig. \ref{fig:3} (a). 
Changing the IDT drive frequency within the excitation bandwidth produces results similar to the one extensively reported at 1 GHz regarding the focusing effect.\\ 
The full SAW excitation bandwidth can be estimated by applying a multi-frequency bias waveform at the IDT and then looking at the Fourier-transform of the time signal from the LDV beam spot placed, for example, close to the focusing region (blue dot - input in Fig. \ref{fig:4} (a)). The displacement amplitude of the SAW excitation is reported in Fig. \ref{fig:4} (b); the measured spectrum is a result of the superposition of three different excitation windows, each of them composed by $n_{BW}=8001$ monochromatic signals, multiplexed to generate the multi-frequency carrier signal. The total, integrated voltage in each window was of 3.5 $V_{RMS}$; to compare the result with the one given by the monochromatic, single frequency 1 GHz tone, one should consider that the signal is normalized considering the voltage power spectral density (VSD), defined as $\lvert V^2\rvert/(2\delta f)$, with $\delta f$ being the frequency measurement resolution (see, for example \cite{Aspelmeyer2014}). Scaling opportunely the voltage in the single frequency measurement for a factor $\sqrt{\delta f}/n_{BW}$, the $\sim$ 300 pm displacement amplitude at 1 GHz reported in Fig. \ref{fig:3} should then translate to roughly 6.5 pm at the same frequency, given a total applied voltage 3.5 $V_{RMS}$. Considering some uncertainty in the spatial position of the laser beam, Fig. \ref{fig:4} (b) well reproduces the expected result, showing a displacement of 5$\pm$1 pm. 
The spectral shape of the signal is the one expected from a SAW IDT excitation, which can be approximated with multiple step-functions in the real space. This produces a sinc-like spectrum, with the main and secondary maxima clearly visible in the figure \cite{Morgan2007}. The spectral shape of the input signal is modified by the propagation through the mechanical nanostring. The output signal can be seen in Fig. \ref{fig:4} (c) ((red dot - output in Fig. \ref{fig:4} (a))). From the reduction of the signal amplitude around the maximum of the spectrum we can extract a total loss of about 5.2 dB. Identifying the contribution from the different loss mechanisms, including propagation loss, coupling loss and parasitic effects due to the frame oscillations, is a challenging task in a short nanostring and will be the object of future investigations. Normalizing the spectral displacement main peaks to one and comparing the input and output signals, it is possible to emphasize the significant difference in the main spectral features (see Fig. \ref{fig:4} (d)). The output spectrum has relatively sharp peaks at 0.985 GHz, 1.001 GHz and 1.018 GHz. 
\begin{figure}[ht!]
\centering
\includegraphics[width=7cm]{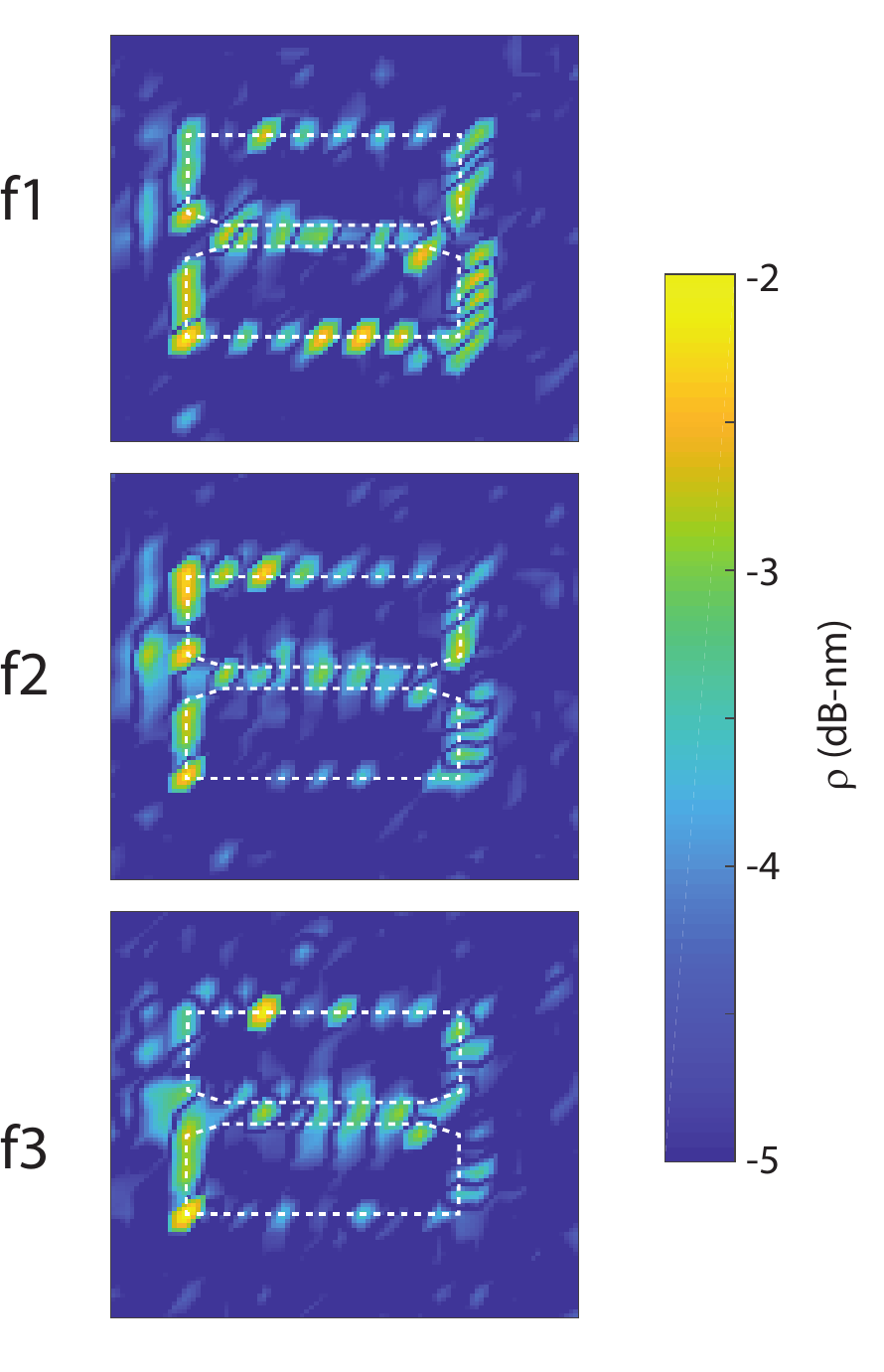}
\caption{Low resolution displacement amplitude maps for the three excitation frequencies reported in Fig. \ref{fig:4}.}
\label{fig:5}
\end{figure}
These peaks are likely arising from the excitation of resonant modes in the mechanical system composed by the nanostring and the released edge of the frame. Indeed, while we did not observe any strong sign of standing waves in the nanostring, which would indicate high-quality factor resonant modes, we expected that our waves have a mix of standing and traveling components. We can easily imagine that, while a strong forward wave is continuously fed by the SAW, only a weak back-scattering component could be created, due to the absence of proper mirrors in the system. Nevertheless, counter propagating waves can still be present due to wave reflection at the interface between AlN+Si and Si regions. This effect complicates the evaluation of the displacement field onto the nanobeam, since the weak standing wave component feels the full nanostring plus frame eigenstate structure. More insights on the impact of the frame can be obtained by performing extended map scans on a wide area around the nanostring region. Figure \ref{fig:5} reports the displacement amplitude maps for the three frequencies indicated in Fig. \ref{fig:4} (c). Although the maps have a low resolution in order to minimize possible phase drifts during the measurements (which nevertheless are still partially present and indicated by the tilted shape of the displacement field lobes), it is possible to identify the role of the hanging frame in the vibration excitation. In all cases the frame is significantly excited with vibration amplitudes equal or even exceeding the ones of the string itself. This complicates the analysis of the system eigenmodes and could somewhat degrades the nanostring potential performance. On the other hand, we cannot exclude that the frame could be beneficial for vibration coupling into the device itself by extending the spatial profile of the mechanical eigenmode and by relaxing the strong focusing requirement for exciting the narrow silicon string.\\
A better assessment of the vibrations injected into the nanostring is performed by switching back to a monochromatic excitation at 2.25 V$_{RMS}$ and 1 GHz frequency. By performing line scans along the nanostring length, we can estimate the displacement induced by the SAW. In addition, by performing parallel line scans on the etched region it can be checked if the SAW propagates underneath the device in the Si substrate. This is expected since the penetration depth of a SAW is of the order of the wavelength (about 4 $\mu m$ in our case), which is larger than the 1 $\mu$m thick SiO$_2$ sacrificial layer, as shown in Fig. \ref{fig:6} (a). A strong excitation on top of the nanostring is accompanied by weaker ones on the bottom of the device window. It is possible to verify that a coherent wave is propagating below the suspended structure by looking at the phase map in a 2D scan measurement similar to the ones reported in Fig. \ref{fig:5}. As can be seen in Fig. \ref{fig:6} (b), a wavefront is propagating on the silicon substrate, from one end of the etched region to the other. A more quantitative assessment of the overall displacement excitation can be done by inspecting the results in Fig. \ref{fig:6} (c), where the measured signals on the nanobeam center and on one of the two sided on the Si substrate are compared. While on the nanostring the displacement amplitude reaches about 1 nm, it is reduced by about one order of magnitude with respect to the substrate underneath. Despite the excitation splitting on top and bottom layers, considering the average vertical displacement on the maxima and an applied voltage of 2.25 V$_{RMS}$, we can estimate an \textit{external} mechanical peak excitation efficiency, $\eta_m=\rho/V_{RMS}$, of about 519 $\pm$ 38 pm/V. The efficiency estimate is obtained considering the vibrations only on the nanostring; the complicated mode structure of our systems and the limited lateral scan accuracy somewhat impact on a more refined evaluation of the excitation efficiency.\\ 
A useful quantity to evaluate is the Displacement Spectral Density (DSD), which is defined as DPSD=$\left|z\right|^2/(2 \delta f)$. The integral of the DSD over a mechanical resonance gives the expectation value of the displacement square, $\langle z^2 \rangle$, even when the systems is actuated through (thermal) noisy excitation \cite{Aspelmeyer2014}. For a resonance with weak damping, it is also possible to theoretically estimate the thermal displacement square, $\langle z_{th}^2 \rangle$, using the equipartition theorem and considering a temperature T \citep{Aspelmeyer2014}:
\begin{equation}\label{eq:equip}
\langle z_{th}^2 \rangle = \frac{k_B T}{m_{eff} \omega_m^2},
\end{equation}
where $k_B$ is the Boltzmann constant, $\omega_m$  resonant angular frequency and $m_{eff}$ effective mass of the resonator. The latter can be evaluated using commercial FEM solvers (COMSOL Multiphysics); in our case we obtained a $m_{eff}$ in a range from 2-5 $\times 10^{-14}$ kg, considering the few modes of the nanostring plus frame system oscillating around $\omega_m/2\pi$ = 1 GHz. Plugging these values in eq. (\ref{eq:equip}), one finds the expected displacement square due to thermal motion at room temperature, $\langle z_{th}^2 \rangle$ = 0.002 - 0.005 pm$^2$, considering the different oscillation modes as before.
A direct measure of the DSD with our coherent technique can give some hints about the coherent displacement induced by the SAW. Fig. \ref{fig:6} (d) reports the average DSD obtained in a different experiment, where we considered the area enclosed by the dashed line in Fig. \ref{fig:6} (d). The device was excited with a 2.25 $V_{RMS}$ monochromatic tone oscillating at 1 GHz. The DSD is plotted along the nanobeam length ($\hat{x}$), where the average has been taken along the nanobeam width direction ($\hat{y}$). The measurement bandwidth was $\delta f$=625 kHz. Considering a single, total DSD value, obtained by averaging along along $\hat{x}$, it is possible to estimate a SAW-driven displacement square $\langle z_{SAW}^2 \rangle$ of about 8 pm$^2$. We can roughly compare the linear displacement driven by the coherent SAW and by thermal motion by considering the ratio of the square of DSD, i.e. $\eta_{coherent}=sqrt{\langle z_{SAW}^2 \rangle}/sqrt{\langle z_{th}^2 \rangle}$. A larger injection of coherent vibrations translates in a $\eta_{coherent}$ in a range from 40 to 65, where the different mechanical modes have been considered. In average, our coherent control of the mechanical motion is about 50 times stronger than the thermal noise, granting a strong and reliable SAW-based operation even when working at room temperature. Of course, our rough estimate suffers from multiple limitations, starting from the assumptions leading to eq. (\ref{eq:equip}), as having a single mechanical mode with low damping; more accurate evaluations will be performed in dedicated sets of samples.\\
\begin{figure}[t!]
\centering
\includegraphics[width=8cm]{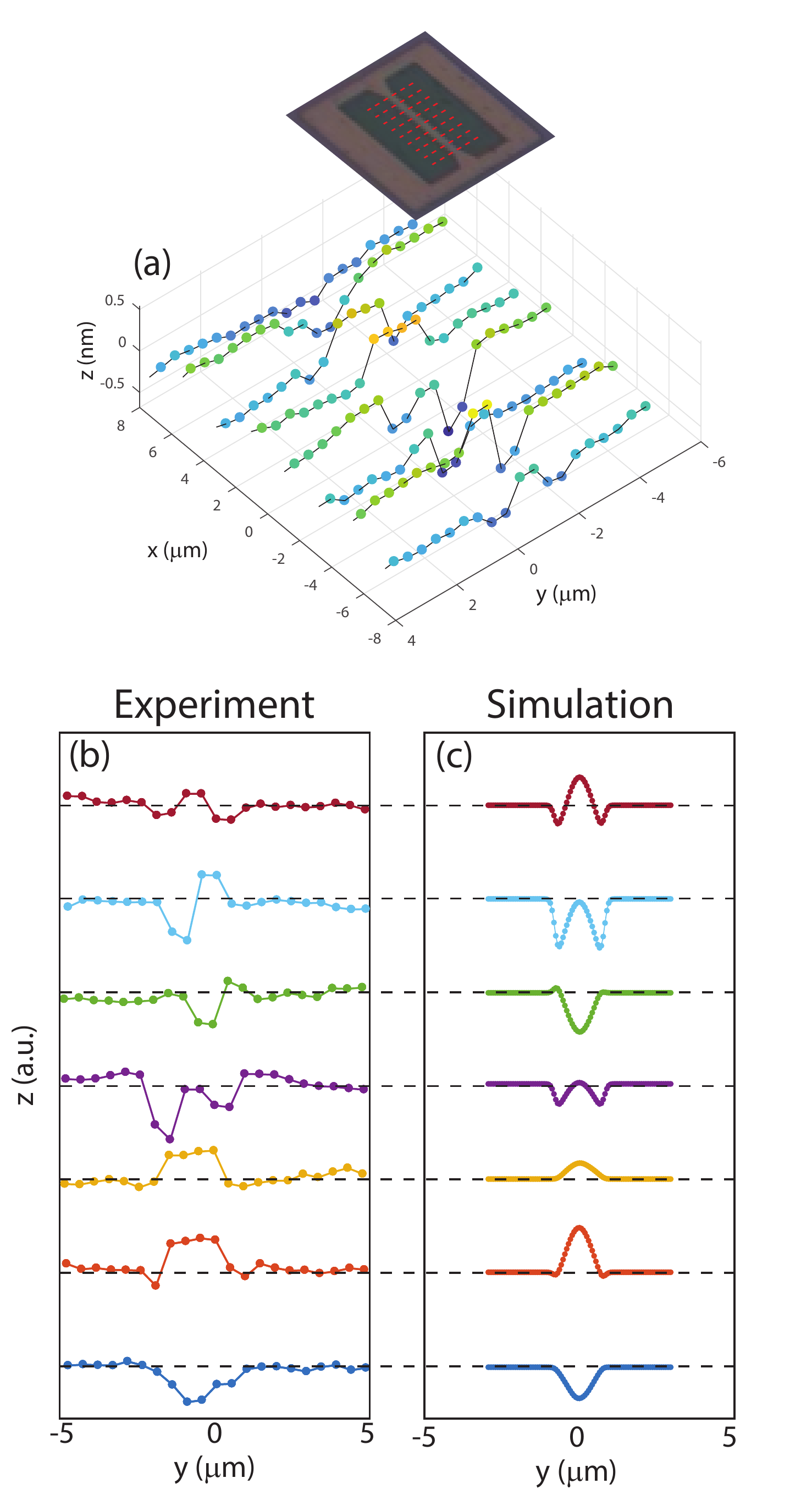}
\caption{(a): 3D sketch of the cross-line scan measurement showing the displacement of the nanobeam excitation. Waterfall plot of experimental (b) and simulated displacement (c) obtained from a time domain simulation.}
\label{fig:7}
\end{figure}

Although a strong vibration amplitude can be reached, the spatial shape of the nanostring displacement of Fig. \ref{fig:6} (c) and (d) do not show a regular, harmonic form, at odds with the weaker substrate signals reported in Fig. \ref{fig:6} (c). This is due to the fact that the nanostring geometry has not been optimized to work around 1 GHz, resulting in excitation of high order modes, further complicated by the presence of the hanging frame which, as we have seen, plays an important role regarding the whole system excitation. Some insights on the mechanical excited modes can be obtained by considering line scans crossing the nanostring along its width. Such a measurement is reported in Fig. \ref{fig:7} (a), along with a visual help of the scan lines superimposed on the optical microscope image. Even if a straight comparison between the measurement obtained and device geometry is not easily done due to the rough resolution of the stage movement, it is possible to observer that few scans have displacement antinodes at the nanobeam edges and nodes at the nanobeam center. Clearly, this is not an optimized configuration when working with a laser spot size of the same order of magnitude of the nanostring lateral width (1.5 $\mu$m). Nevertheless, the experimental data can be tentatively compared with finite-element method simulations (COMSOL Multiphysics) of the nanostring plus hanging frame geometry. Nanocrystalline Si has been simulated using a 157 GPa Young's modulus and 0.22 Poisson's ratio, which are compatible with values reported in the literature \cite{Lu2012}. In order to reproduce our hybrid standing/traveling wave system, we run time-domain simulations, in which a 1 GHz mechanical source at one of the nanostring edges is slowly turned on; after a transient phase, we can then estimate the displacement field propagating into the nanostring. A qualitative comparison of the experimental data and simulations can be done by inspecting the waterfall plot of panel (b)-(c) in Fig. \ref{fig:7}. 
Here the phase of the measurement and simulation time have been chosen to optimize a qualitative data comparison. Moreover, the simulation data has been post-processed by considering a convolution with a Gaussian function of 300 nm width, roughly representing the effect of a finite laser beam profile. Even if the spatial resolution required for a more quantitative comparison is beyond our stage movement capabilities, we can find a good qualitative agreement in most of the reported linescans, suggesting that we are experimentally observing the simulated mode.

\section{Conclusions}

In this manuscript we have shown a $\sim$1 GHz electromechanical system where a Si nanostring is excited through SAW generated in a piezoelectric AlN layer embedded in the chip. Our hybrid system allows the use of high-frequency SAW with silicon, which has unrivaled characteristics for electronics and photonics technologies. Our approach combines the expertise of fabrication of Si-based devices and the versatility of SAW, which are already making some of the best high-frequency filters overall and are slowly entering the world of quantum applications \cite{Delsing2019}. The AlN/Si platform is completely silicon compatible and competes favorably with other monolithic and hybrid approaches showing an external mechanical excitation efficiency of 519 pm/V, using optimized IDT with a simple, not-optimized Si nanostring, making it feasible to reach very strong mechanical drives with applied biases of just few volts. This makes the system appealing for low power, high frequency optical modulation and RF filtering. Furthermore, the device characterization has been performed in ambient conditions (room temperature and atmospheric pressure), showing the robustness of our technology against thermal noise and pressure-induced damping. Future device realizations will see the use of proper optomechanicals crystal replacing the nanostring, with the added value of largely increased sensitivity to mechanical displacement as well as the possibility to coherently drive the optomechanical crystal for highly efficient, unidirectional wavelength conversion from microwaves to optical regime. 

\section{Acknowledgments}

The authors wish to thank M. Cecchini for access to the LDV. This work was supported by the FET-Open PHENOMEN project (GA 713450). DNU gratefully acknowledges funding from a Ramón y Cajal postdoctoral fellowship (RYC-2014-15392). MFC acknowledges a PhD studentship granted by the Spanish S. Ochoa project (AEI, grant no. SEV-2017-0706).

\bibliography{Pitanti_AlN_Si_nanostring}% Produces the bibliography via BibTeX.

\end{document}